\documentclass[prd,epfs,epsfig,aps]{revtex4}

\usepackage{graphicx}%
\usepackage{dcolumn}
\usepackage{amsmath}

\begin{document}

\title{Galerkin Method in the Gravitational Collapse: a Dynamical System Approach}

\author{H. P. De Oliveira}
 \email{Henrique@Fnal.Gov}
\affiliation{{Nasa/Fermilab Astrophysics Center \\
Fermi National Accelerator Laboratory, Batavia, Illinois, 60510-500.}\\
And \\
{\it Universidade Do Estado Do Rio De Janeiro }\\
{\it Instituto De F\'{\i}sica - Departamento De F\'{\i}sica Te\'orica,}\\
{\it Cep 20550-013 Rio De Janeiro, RJ, Brazil}}

\author{I. Dami\~ao Soares}
\email{ivano@cbpf.br}
\affiliation{{\it Centro Brasileiro de Pesquisas F\'\i sicas}\\ {\it  Rua Dr.
Xavier Sigaud, 150} \\ {\it CEP 22290-180, Rio de Janeiro, RJ, Brazil}}

\date{\today}





\date{\today}


\begin{abstract} 
We study the general dynamics of the spherically symmetric gravitational
collapse of a massless scalar field. We apply the Galerkin projection method
to transform a system of partial differential equations into a set of ordinary
differential equations for modal coefficients, after a convenient truncation
procedure, largely applied to problems of turbulence. In
the present case, we have generated a finite dynamical system that reproduces
the essential features of the dynamics of the gravitational collapse, even for
a lower order of truncation. Each initial condition in the space of modal
coefficients corresponds to a well definite spatial distribution of scalar
field. Numerical experiments with the dynamical system show that depending on
the strength of the scalar field packet, the formation of black-holes or the
dispersion of the scalar field leaving behind flat spacetime are the two main
outcomes. We also found numerical evidence that between both asymptotic states,
there is a critical solution represented by a limit cycle in the modal space
with period $\Delta u \approx 3.55$.
\end{abstract}

\maketitle

\newpage

Gravitational collapse is one of the most fascinating problems studied in
General Relativity. Remarkable works helped the understanding
of the process of  gravitational collapse\cite{review}, but there are still
crucial open questions concerning, for instance, the Cosmic Censorship
Conjecture, and more recently the existence of critical phenomena in
gravitational collapse. We are going to focus our attention here to this last
issue. 

The pioneering analytical results of Christodoulou\cite{chris} together with
numerical work\cite{gold} about the
collapse of massless scalar fields basically established that, depending on the
strength of the scalar field packet, two types of outcomes are expected to take
place: the scalar field disperses to infinity leaving behind flat spacetime
after an initial stage of collapse, or the scalar field collapses to form a
black 
hole. Later Choptuik\cite{chop}, abandoning the restriction to almost 
stationary initial data, performed definite numerical work establishing the
idea of critical behavior in the gravitational collapse and of a threshold of
black hole formation. Originally, his numerical study was realized in the
collapse of spherically symmetric distributions of massless scalar fields, and
later several other fields such as perfect fluid\cite{pf}, Yang-Mills\cite{ym},
charged and complex scalar fields\cite{ccsf}, massive scalar
fields\cite{massive}, gravitational waves\cite{gw}, etc\cite{gundlach}, were
taken into consideration. Critical phenomena emerge as a robust feature in
gravitational collapse, and might be attributed to the nonlinear nature of the
Einstein field equations.

Here we present some interesting preliminary results concerning the
application, for the first time, of the Galerkin projection
method\cite{galerkin} in the gravitational collapse of a massless scalar
field. The idea is to treat the general gravitational collapse of a spherically
symmetric massless scalar field without imposing any additional symmetry like
some kind of self-similarity. The Galerkin method is an approximative technique
that allows to transform a finite set of partial differential equations into an
infinite countable set of coupled ordinary differential equations associated to
the infinite modes of Fourier-type in which the solutions are decomposed.
Truncation of these equations to a certain order of modes may give, under
certain conditions, a finite dimensional dynamical system which approximates
fairly well the exact dynamics. The application of this method  in problems of
fluid mechanics has brought interesting new results that has shed light in the
understanding of turbulence, where the most famous application of the Galerkin
method is the Lorenz system \cite{lorenz} (resulting from a truncation to third
order of the Galerkin projection of Navier-Stokes equation) that describes the
chaotic Rayleigh-Benard convection.

Basically the Galerkin projection method consists in expanding a set of the
system variables in a series of orthogonal polynomials (chosen as basis of the
projection space $\textit{X}$), substitution in the dynamical equations and
obtaining a countable infinite set of ordinary differential equations for the
{\it modal} coefficients. In order to carry out the Galerkin projection, the
space $\textit{X}$ must be an inner product space and the basis functions must
satisfy the boundary conditions implicit in the definition of $\textit{X}$ such
that any candidate to a solution automatically satisfies the boundary
conditions. Truncation of the system by setting $\textit{modal}$ terms beyond a
certain order equal to zero (Galerkin approximation) results in a finite
dimensional system which $\textit{presumably}$ yields an adequate approximation
to the infinite dimensional dynamics if the truncation is at sufficiently high
order. To gain insight in the types of dynamics that are possible we will
consider here a low order truncation; while our truncation is not of high
enough order to model the real behavior of our system in a completely faithful
way, the resulting solutions give an indication of the type of qualitative
behavior of which the actual system is capable.


We consider the general spherically symmetric line element given by 

\begin{equation}
d s^2 = -g \bar{g}\,d u^2 - 2 g\,d u\,d r + r^2\,d \Omega^2, \label{eq1}
\end{equation} 

\noindent where $g=g(u,r)$, $\bar{g}=\bar{g}(u,r)$ and $d \Omega^2 = d
\theta^2 + sin \theta^2\,d \varphi^2$; $u$ is the usual retarded null
coordinate and $r$ a radial coordinate which measures the proper area of the
two sphere $d \Omega^2$. The massless scalar field $\phi(u,r)$ is the only
source of curvature, such that the relevant field equations that describe the
dynamics are

\begin{eqnarray}
(ln\,g)_{,r}&=& \frac{r}{2}\,\phi_{,r}^2  \label{eq2}\\
(r \bar{g})_{,r}&=&g  \label{eq3}\\
r \phi_{,ur} + \phi_{,u} - \bar{g} \left(\phi_{,r} + \frac{r}{2}
\phi_{,rr} \right) &=& \frac{r}{2} \bar{g}_{,r} \phi_{,r}.
\label{eq4} 
\end{eqnarray}

%

\noindent Eqs. (\ref{eq2}) and (\ref{eq3}) follow directly from the
Einstein field equations, and the dynamics is completely encompassed by the
Klein-Gordon equation (\ref{eq4}). Another important quantity to be introduced
is the mass function, $m(u,r)$,

\begin{equation}
1 - \frac{2 m(u,r)}{r} \equiv g^{\mu \nu}\,r_{,\mu} r_{,\nu} =
\frac{\bar{g}}{g}. \label{eq5}
\end{equation}

\noindent This quantity is interpreted as the effective gravitational mass
inside the 2-sphere of radius $r$, and agrees with the Bondi and ADM masses in
the asymptotic flat spacetimes.

As we have discussed briefly, the boundary conditions are of fundamental
importance for the Galerkin method, since they dictate the convenient set of
basis functions. We then assume that the scalar field must satisfy  the following
boundary conditions, $\phi(u,0) = 0$ and $\phi(u,\infty) = 0$. Regularity of
the spacetime at the origin requires $\bar{g}(u,0) = 1$, while the choice of
$u$ as the proper time at the origin fixes $g(u,0) = 1$.

After introducing the basic equations together with the appropriate boundary
conditions, we are ready to apply the Galerkin method. For the sake of
convenience, we introduce an alternative radial coordinate $\xi$ related to
$r$ by $\xi = ln r$, where this new radial coordinate varies from $-\infty$ to
$+\infty$ corresponding, respectively, to $r = 0$ and $r = \infty$. Then, the
chosen orthogonal set of basis functions for the projection space $X$ is   

\begin{equation}
\psi_k(\xi) = e^{-\xi^2/2} H_k(\xi),  \label{eq6}
\end{equation}

\noindent where $H_k(\xi)$ are the Hermite polynomials. The orthogonality is
defined by the inner product 

\begin{equation}
\left<\psi_k(\xi),\psi_j(\xi)\right> \equiv \int^{+\infty}_{-\infty} \psi_k(\xi)
\psi_j(\xi)\,d \xi = 2^k \sqrt{\pi} k!\,\delta_{k j}.  \label{eq7}
\end{equation}

The following decomposition for the scalar field  $\phi(u,\xi)$ is
proposed 

\begin{equation}
\phi(u,\xi) = \sum_{k=0}^{N} \phi_k(u) \psi_k(\xi),\;\;\;
\label{eq8} 
\end{equation}

\noindent where $\phi_k(u)$ are the modal (Fourier-type) coefficients and $N$ indicates the
order of truncation. We remark that the precision of the above decomposition, in
the sense of approaching the actual solution, depends directly on $N$. The
boundary conditions for $\phi$ are automatically  satisfied since
$\psi_k(-\infty) = \psi_k(+\infty) =0$ for any $k$. 

The decomposition of $\phi(u,\xi)$ will be considered as the fundamental
piece of our strategy of applying the Galerkin method to the system of
Eqs. (\ref{eq2})-(\ref{eq4}). Before going through the Klein-Gordon equation,
Eq. (\ref{eq4}), we need to express the metric functions $g(u,\xi)$ and
$\bar{g}(u,\xi)$ conveniently in terms of the basis functions. Thus,
considering Eq. (\ref{eq2}) with the  new spatial variable $\xi$, and
introducing the decomposition of the scalar field given by Eq. (\ref{eq8}), we
have 

\begin{equation}
g(u,\xi) = exp\left(\frac{1}{2}\,\int^{\xi}_{-\infty} \sum_{k=0}^N
\sum_{j=0}^N \phi_k(u) \phi_j(u) \frac{d \psi_k}{d \xi^{\prime}} \frac{d
\psi_j}{d \xi^{\prime}}\,d \xi^{\prime}\right). \label{eq9}
\end{equation}    

\noindent The integration can be performed without difficulty for any order of
the truncation; also from Eq. (\ref{eq9}) we have that $g(u,\xi)$ satisfies the boundary
condition $g=1$ at the origin $\xi=-\infty$, for all $u$. 

Now, the next step is to consider Eq. (\ref{eq3}) that relates the metric
functions $\bar{g}(u,\xi)$ and $g(u,\xi)$. This equation can be written as

\begin{eqnarray}
\bar{g}(u,\xi) = e^{-\xi} \int^{\xi}_{-\infty} e^{\xi^{\prime}}
g(u,\xi^{\prime})\,d \xi^{\prime} = g(u,\xi) - \frac{\partial 
g}{\partial \xi} + \frac{\partial^2 g}{\partial \xi^2} 
- e^{-\xi}\int^{\xi}_{-\infty} e^{\xi^{\prime}} \frac{\partial^3 g}{\partial
\xi^{\prime^3}}\,d \xi^{\prime}, \label{eq10}  
\end{eqnarray} 

\noindent where the term on the right-hand side  arises from successive
integration by parts. At this point, $\bar{g}(u,\xi)$ can be approximated in
the following way

\begin{equation}
\bar{g}(u,\xi) \simeq g(u,\xi) + \sum^J_{k=1} (-1)^k\,\frac{\partial^k
g}{\partial \xi^k}. \label{eq11}
\end{equation} 

\noindent Basically this approximation will be used in order to compute
$\bar{g}$ from $g$, namely, as an integral of Eq. (\ref{eq10}).

The last step of the Galerkin method is to substitute Eqs. (\ref{eq8}), (\ref{eq9})
and (\ref{eq11}) into the Klein-Gordon equation that dictates the dynamics of the
scalar field. The dynamical system for the modal coefficients $\phi_k(u)$ is
obtained after projecting the Klein-Gordon equation into the $nth$ mode
$\psi_n(\xi)$, $n=0,1,2,..,N$, or 

\begin{eqnarray}
& & \sum_{k=0}^N \dot{\phi}_k(u) \left<e^{\xi}
\left(\frac{d \psi_k}{d \xi}+\psi_k\right),\psi_n(\xi)\right> - \frac{1}{2}
\sum^{N}_{k=0} \phi_k(u) \left<\bar{g}(u,\xi) \left(\frac{d \psi_k}{d
\xi}+\frac{d^2 \psi_{k}}{d \xi^2}\right),\psi_n(\xi)\right> =
\nonumber  \\  
& & \frac{1}{2} \sum_{k=0}^{N} \phi_k(u) \left<\frac{\partial
\bar{g}}{\partial \xi} \frac{d \psi_{k}}{d
\xi},\psi_{n}(\xi)\right>,\;\;\; n=0,1,2..,N \label{eq12} 
\end{eqnarray}

\noindent where dot stands for derivative with respect to $u$. The result is a
set of equations of the type 

\begin{equation}
\dot{\phi}_k(u) = F_k(\phi_j),\;\; k=0,1,2...N \label{eq13}
\end{equation}

\noindent where the number of equations is defined by the order $N$ of
truncation. The expressions on the right-hand-side of Eqs. (\ref{eq13}) are
much involved, even for low $N$, and will not be given here. Once the modal
coefficients are known either analytically or numerically, the overall
spatio-temporal behavior of the scalar field and the metric functions are
determined.

Before the discussion of the qualitative aspects of the dynamical system, it
will be very useful to establish a convenient way to choose the initial
conditions. In our approach, a given set of initial conditions  $(\phi_0(0),
\phi_1(0),.., \phi_N(0))$ corresponds to the initial spatial distribution
of scalar field given by $\phi(0,\xi) = \sum^N_{k=0} \phi_k(0)
\psi_k(\xi)$. Thus, let us consider a Gaussian initial data

\begin{equation}
\phi(0,\xi) = A e^{2 \xi}\,e^{-(e^{\xi}-0.2)^2},
\label{eq14} 
\end{equation}

\noindent where $A$ is the amplitude of the distribution. The above expression
can be decomposed with respect to the basis functions $\psi_k(\xi)$, 
determining the set of initial conditions, since

\begin{equation}
\phi_k(0) =
\frac{\left<\phi(0,\xi),\psi_k\right>}{\left<\psi_k,\psi_k\right>} =
\frac{1}{2^k \sqrt{\pi} k!}\,\int^{\infty}_{-\infty}\phi(0,\xi) \psi_k(\xi) 
d\,\xi,\;\;k=0,1..,N,  \label{eq15}
\end{equation}

\noindent such that the information about the initial strength of the scalar
field is contained in the initial set $(\phi_0(0),\phi_1(0),.., \phi_N(0))$,
which depends directly on the amplitude $A$ of the initial distribution of the
scalar field. As illustration, for the value $A=2$ of the amplitude of the
initial distribution we obtain the initial conditions of the modal coefficients
for the projections up to order $N=3$, $\phi_0(0)=0.6782306563805649,
\phi_1(0)=-0.05284561865739180, \phi_2(0)=-0.09123038744236496,
\phi_3(0)=-0.0007815958656455472$. We note that the higher the order $N$ of
truncation, the more the corresponding modal coefficients decrease; for
instance, if we consider $N=9$, $\phi_9(0) \sim 10^{-6}$. A comparison of the
original function (\ref{eq14}) with the initial condition function constructed
from the partial expansion of order $N=3$ with the modal coefficients given
above, shows clearly that the relevant contributions actually come from the
lower order modes so that, in this aspect, our truncation to order $N=3$
appears sufficient to grasp the essential features of initial scalar function
wave packet.

Now, we are ready to present the basic qualitative aspects of the dynamics in
the space of variables $\phi_k(u)$. The origin is the unique critical
point identified as the Minkowski or flat spacetime, because $\phi(u,\xi)=0$
and $g(u,\xi) = \bar{g}(u,\xi)=1$. The linear analysis about the origin reveals
that it is an attractor for any order $N$ of the truncation. The eigenvalues
associated with the linearized system about the origin are in general complex,
conjugated in pairs, with negative real parts. From the physical point of view,
this characteristic is reasonable due to the fact that small modal coefficients
$\phi_k(0)$ are equivalent to a small amplitude $A$ of the initial data. As a
matter of fact, the scalar field has not enough strength to maintain the
collapse and it eventually disperses leaving behind flat spacetime. This
behavior extends to a nonlinear neighborhood of the origin, generated from
initial modal coefficients $\phi_k(0)$ associated with a finite (but not
sufficiently large) value of the amplitude $A$. This behavior is illustrated in
Figs. 1 for $A=2.0798$, and is denoted {\it subcritical}. Here and in the
remaining of the paper we are considering the Galerkin decomposition with
$N=3$, meaning the presence of four modal coefficients. Fig. 1(a) shows the
evolution of the modal coefficient $\phi_0(u)$; it oscillates for a certain
range of $u$ and then decays to zero. Other modes have a similar behavior. With
the present initial distribution all modes tend to zero, namely, to the origin
of modal space that represents flat spacetime (indeed the origin is an
attractor for any order of truncation). Figs. 1(b,c) depict, respectively, the
scalar field $\phi$ given by (\ref{eq8}) for $N=3$ and the mass function
$m(u,\xi)$ defined from (\ref{eq5}), as functions of $\xi$ for several values
of increasing $u$. Initially  $\phi$ and $m$ may increase but as expected, for
large $u$, both tend to zero, showing that the scalar field has not enough
strength to collapse resulting in a final configuration corresponding to flat
spacetime. The denomination subcritical given above is now clear, because of
its analogy with similar behavior in the literature, in spite of our truncation
at $N=3$. Our computational resources allowed us to go to a truncation of order
$N=5$, with analogous results, suggesting that even a lower order truncation
may grasp the skeleton of the full dynamics.

A second possible configuration, denoted {\it supercritical}, corresponds to
initial conditions connected to a distribution of scalar field with enough
strength to hold the collapse until the formation of a black hole. The scalar
field distribution in this case has amplitude $A$ large than a critical value
$A^*$ to be discussed later. The final outcome is the infinity region of the
modal space characterized by $\sum^N_{k=0} \phi_k^2 = \infty$. In approaching
this region the scalar  field increases as well the metric functions $g$ and
$\bar{g}$. Figs. 2 illustrate this behavior. Fig. 2(a) shows the modal
coefficient $\phi_0(u)$ for the supercritical solution with $A=2.0799$.
After oscillating this mode increases in a very fast way. All remaining modes
$\phi_1(u)$, $\phi_2(u)$ and $\phi_3(u)$ present an identical behavior. It can
happen that at some point $\bar{g}/g \rightarrow 0$, signalizing the formation
of an apparent horizon. At this point the mass function is given by
$m_{ah}=e^{\xi_{ah}}/2=r_{ah}/2$, with the subscript \textit{ah} denoting the
apparent horizon. Fig. 2(b) shows the expression $2 m/r$ plotted as a function
of $\xi$ for several values of increasing $u$. For $u_{ah} \approx 28.34$, $2
m/r \approx 1$ at $\xi_{ah} \approx -0.54$. This indicates the formation of a
black hole. We remark that these numerical values are an
approximation of those of the full dynamics due to the $N=3$ truncation.

So far the results are in agreement with previous numerical
works\cite{chop,massive}. Following the terminology introduced by Choptuik, we
called  subcritical the first class of solutions, whereas the second class is
known by supercritical. Between the two classes of solutions, there is the {\it
critical solution}. The critical solution has an interesting symmetry known as
discrete self-similarity, which is manifested by the periodicity of the metric
and scalar field with respect to spatio-temporal scales. Gundlach and Koike et
al\cite{gundlach,pf,gund2} engendered a very interesting picture in phase space
for which the critical solution is represented by a limit cycle. In our
dynamical system approach derived from the Galerkin method, we obtained
evidence of a limit cycle in the space of modal coefficients, which corresponds
to the critical solution constituting a limiting configuration between the
subcritical and supercritical classes, according to the results of Choptuik. We
proceed now to discuss our result. Consider $A^*$ the amplitude of the scalar
field distribution associated with the critical solution. We have found that as
$A \rightarrow A^*$ the modal coefficients oscillate for some time before
escaping to infinity (black hole formation) or tending to the origin (flat
spacetime). Numerically we can successively adjust $A$ such that we approach
$A^*$ from the region of initial data corresponding to black hole final state,
or from the region corresponding to a flat spacetime final state. The more
precise $A$ is adjusted to approach the actual critical value $A^*$, the
longer is the time the modal coefficients oscillate periodically. Fig. 3(a)
illustrates this behavior. The modal coefficients correspond to the choice
$A=2.079856788846789$, and it can be noted that after $u \approx 50$ the
oscillations are performed steadily with a definite period $\Delta u \approx
3.55$. The Figure shows modes up to $N=3$ but we have checked that all other
modes, up to $N=5$, exhibit a similar behavior also with $\Delta u \approx
3.55$, indicating that we are approaching a periodic orbit in the complete
modal space, or a {\it limit cycle}, which is a representation in the modal
space of the critical solution. Fig. 4 shows the projection of the previous
solution in the 3-dim subspace spanned by the modes ($\phi_0,\phi_1,\phi_2$).
It is a remarkable fact that the presence of the limit cycle in the modal
dynamical space manifests itself in any order of the truncation. This gives a
strong indication that our dynamical system approach to the gravitational
collapse, via the Galerkin projection method, exhibits the basic skeleton of
the dynamics, as that which was found in numerical simulation of
Einstein-Klein-Gordon system. We are presently making a careful evaluation of
the order of the error between a truncation of order $N$ and orders
$N+1$, $N+2$, and so on. The indication is that the error tends to decrease
fast as $N$ increases, suggesting a faithful approach of the dynamics of
the system.

We must finally remark the potentiality of the Galerkin method applied in the
treatment of the dynamics of the gravitational collapse, as a reduction of
the dynamics to a dynamical system problem, which may avoid
extremely elaborated numerical techniques and, in its simplicity, guarantees
that the resulting dynamics is a skeleton of the underlying physics and not a
consequence of numerical artifacts. We are also presently applying the same
method to the case of massive scalar field.

The authors acknowledge the financial support of CNPq.

\newpage

\begin{figure}[ht]
\rotatebox{270}{\includegraphics*[scale=0.5]{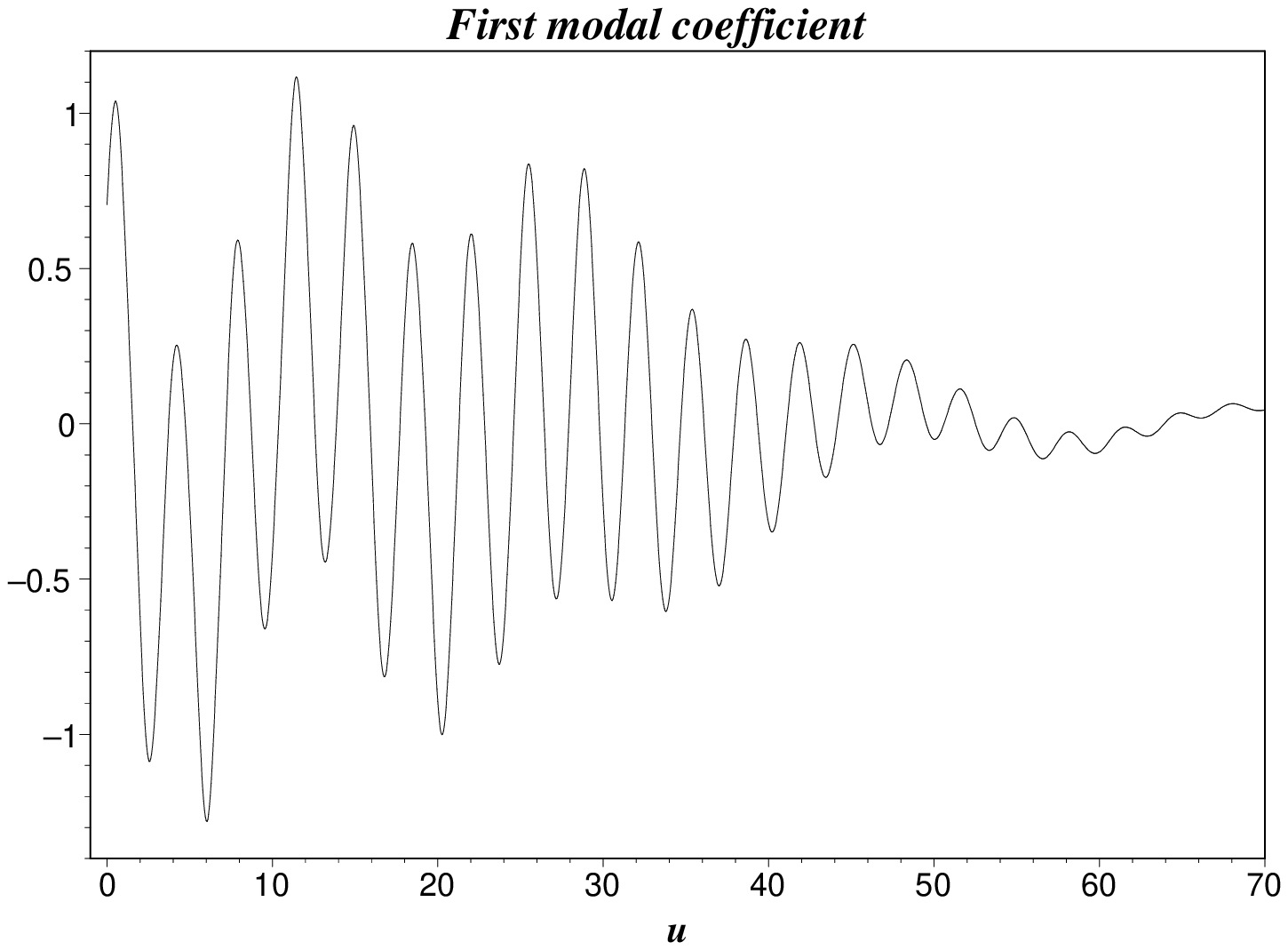}}
\centerline{(a)}
\vspace{0.5cm}
\rotatebox{270}{\includegraphics*[scale=0.5]{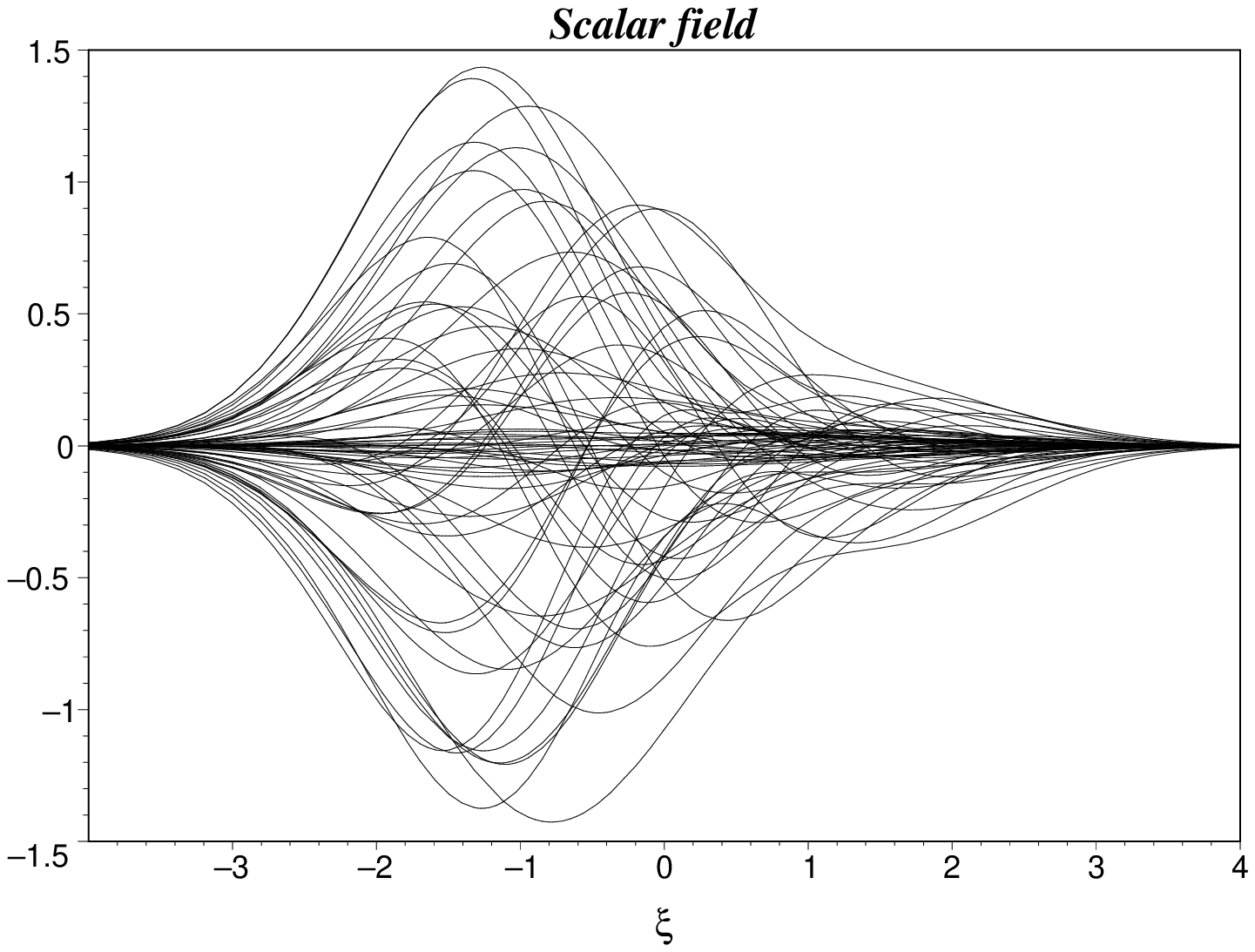}}
\centerline{(b)}
\vspace{0.5cm}
\rotatebox{270}{\includegraphics*[scale=0.5]{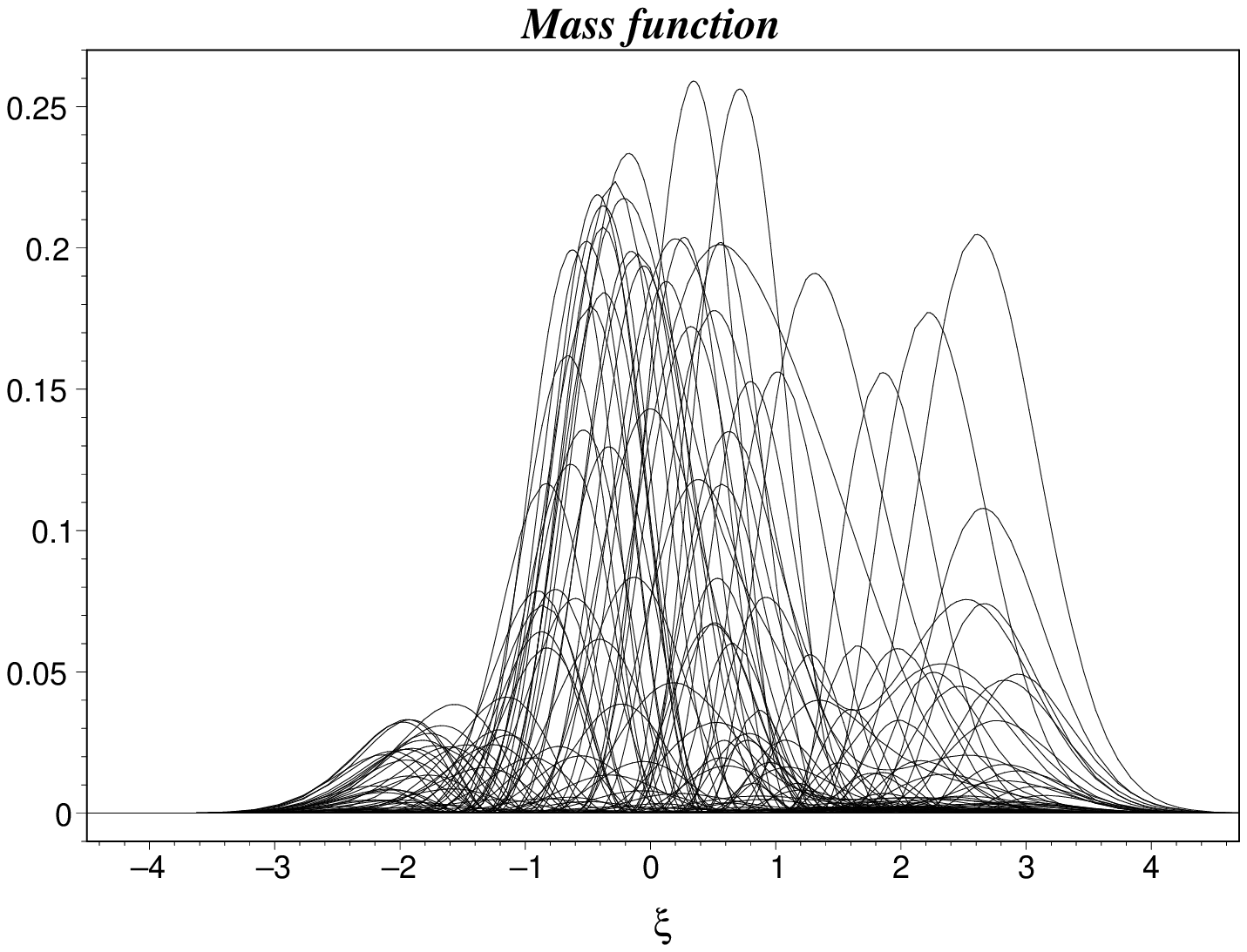}}
\centerline{(c)}
\vspace{0.5cm}
\caption{(a) Evolution of the modal coefficient $\phi_0(u)$ for the
subcritical configuration, which corresponds to the a scalar field
distribution with initial amplitude $A=2.0798$. The remaining modes
have a similar behavior. (b) Behavior of the scalar field as a
function of $\xi$ for several values of increasing $u$. The amplitude of
$\phi(\xi,u)$ increases indicating the initial phase of collapse,
but for large $u$ the scalar field tends to zero corresponding to a final flat
configuration. (c) Distributions of the mass-function $m(\xi,u)$ for increasing
$u$.} \end{figure}

\begin{figure}[ht]
\rotatebox{270}{\includegraphics*[scale=0.5]{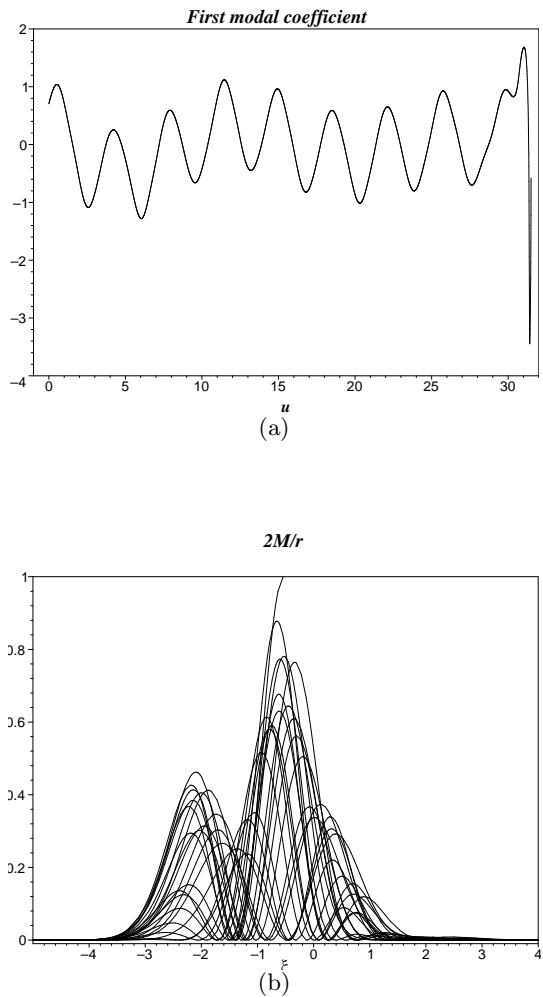}}
\vspace{0.9cm}
\centerline{(a)}
\rotatebox{270}{\includegraphics*[scale=0.5]{fig2b.ps}}
\vspace{0.9cm}
\centerline{(b)}
\caption{(a) Evolution of the modal coefficient $\phi_0(u)$ for the
subcritical evolution with $A=2.0799$. After an initial phase of
oscillations, this mode increases in a very fast way. Other modes
$\phi_{1}$, $\phi_{2}$ and $\phi_{3}$ exhibit identical behavior. (b) $2m/r$
plotted as a function of $\xi$ for several values of increasing $u$. For this
supercritical configuration, $2m/r \approx 1$ indicates the formation of an
apparent horizon at $(\xi_{ah},u_{ah})$ and consequently a black hole.}
\end{figure}

\begin{figure}[ht]
\rotatebox{270}{\includegraphics*[scale=0.5]{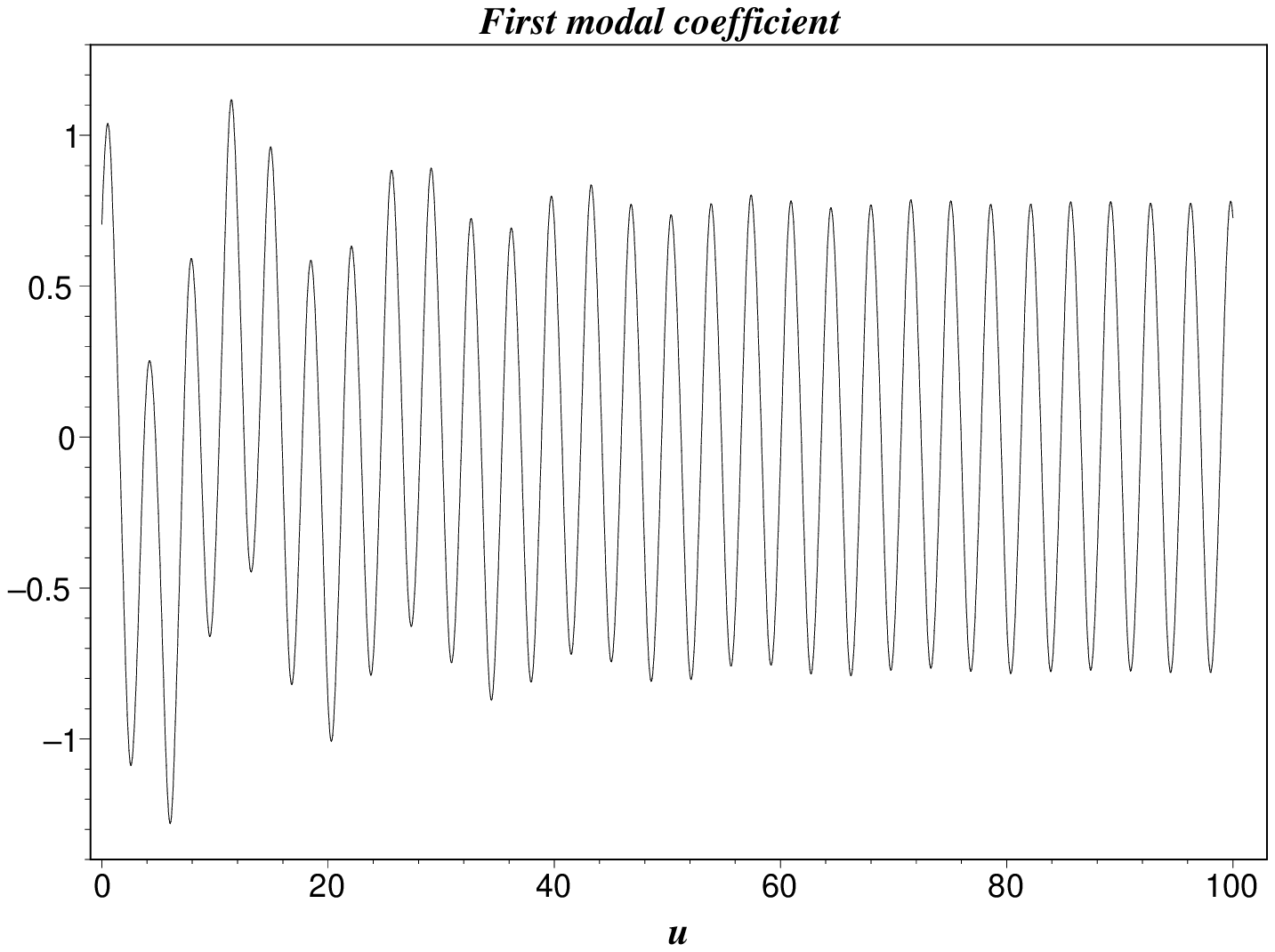}}
\rotatebox{270}{\includegraphics*[scale=0.5]{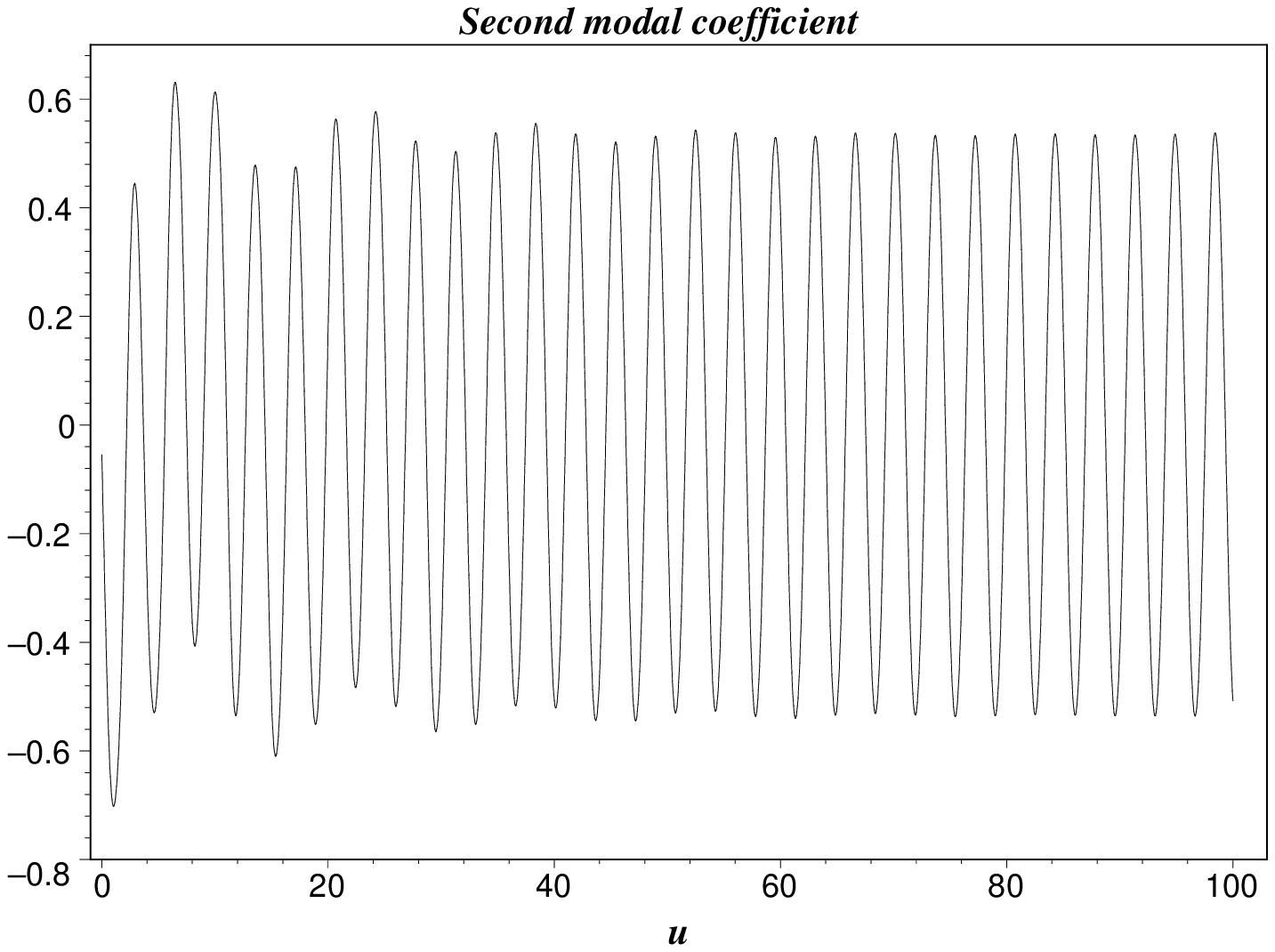}}
\rotatebox{270}{\includegraphics*[scale=0.5]{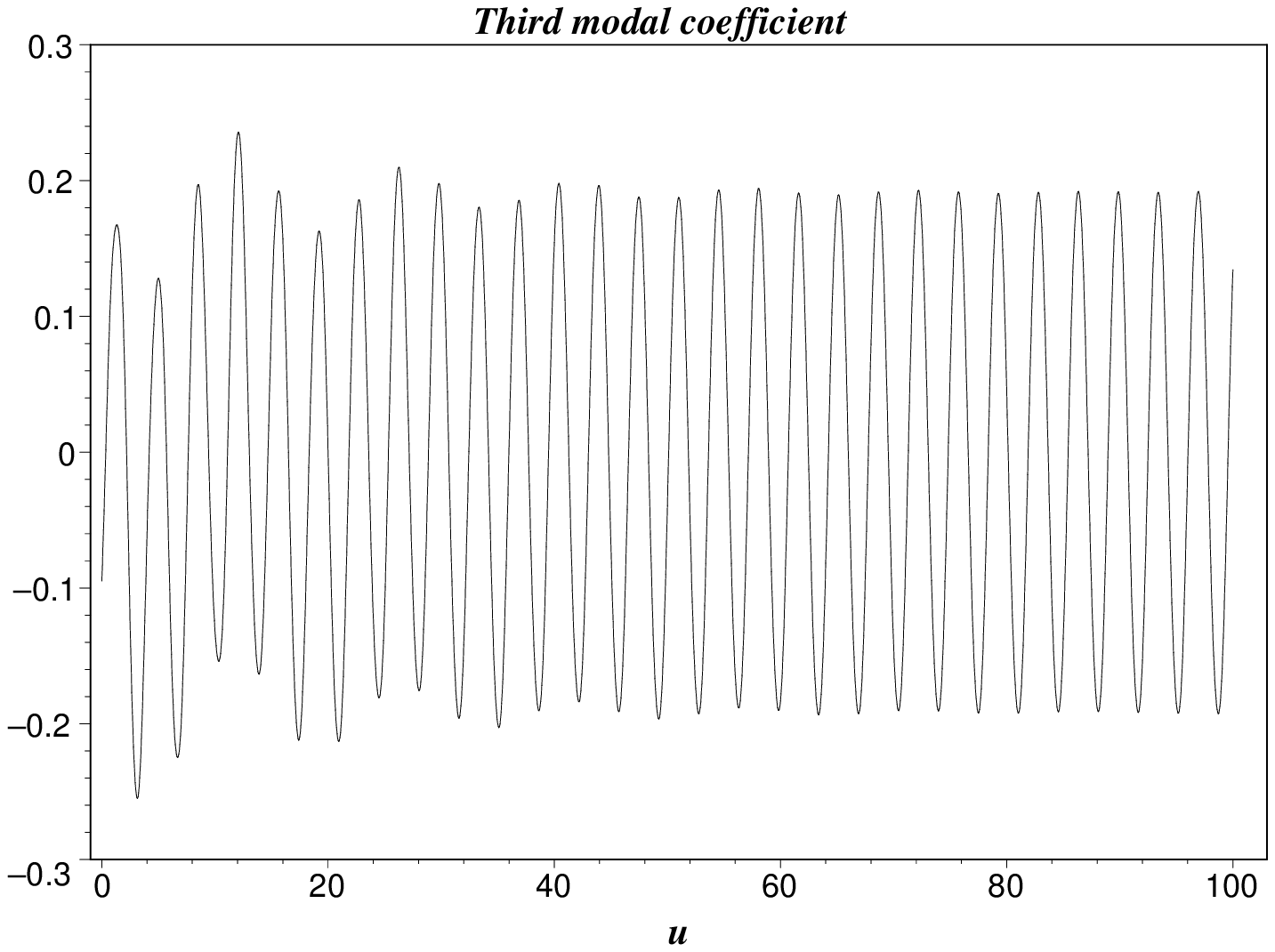}}
\rotatebox{270}{\includegraphics*[scale=0.5]{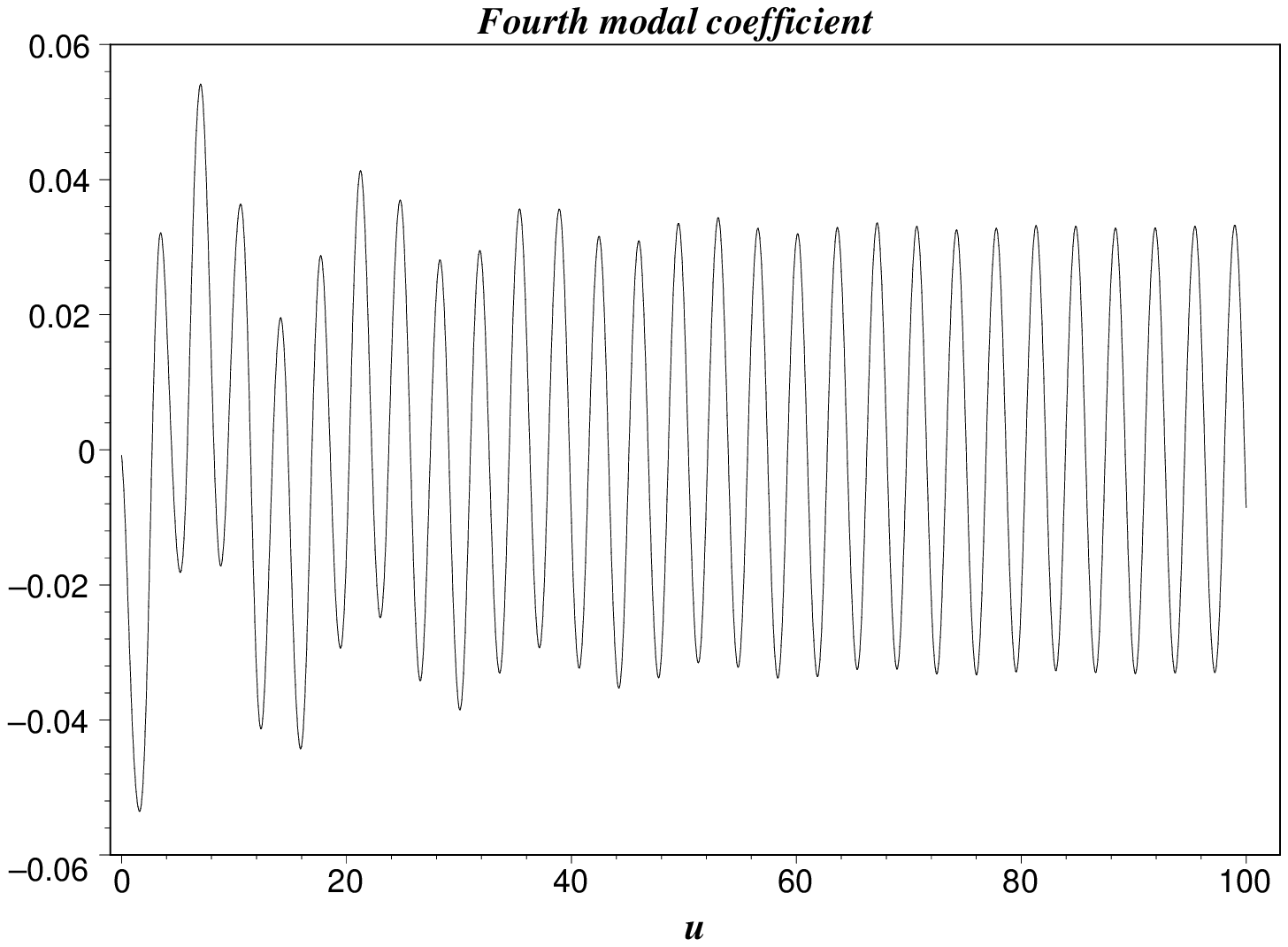}}
\caption{Evolution of the modal coefficients $\phi_0$,
$\phi_1$, $\phi_2$ and $\phi_3$ corresponding to the critical solution
obtained with the choice $A==2.079856788846789$. Note that after $u \approx 50$ the
oscillations are performed steadily with period $\Delta u \approx
3.55$. By observing the amplitudes of the periodic motion along each
modal coefficient, it is clear that they are smaller for higher order modal
coefficients; for instance, for the first modal coefficient the amplitude is
approximately 0.8, and of order of $10^{-2}$ for the last modal coefficient
used in our approximation. Eventually, if more terms were considered in the
decomposition (8), it would be expected smaller corresponding amplitudes of the
periodic motion of such modal coefficients.} \end{figure}

\begin{figure}[ht]
\rotatebox{270}{\includegraphics*[scale=0.5]{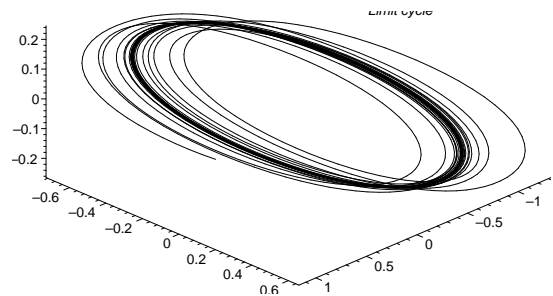}}
\caption{The limit cycle present in the space of modal space for our
case of truncation $N=3$ projected into the 3 dim subspace spanned by
the modes ($\phi_0$,$\phi_1$,$\phi_2$).}
\end{figure}

\end{document}